\documentclass[journal]{IEEEtran}

\ifCLASSINFOpdf
\else
\fi

\usepackage{amssymb}
\usepackage{amsmath}
\usepackage{epsfig}
\usepackage[caption=false,font=footnotesize]{subfig}
\usepackage{cite}

\begin{document}
\title{Characterization of Broadband Focusing Microwave Metasurfaces at Oblique Incidence}

%
%
\author{Ashif A. Fathnan, Toufiq M. Hossain, Dadin Mahmudin, Yusuf N. Wijayanto,~\IEEEmembership{Member,~IEEE}
        \\ and~David A. Powell,~\IEEEmembership{Senior~Member,~IEEE}
\thanks{Ashif A. Fathnan, Toufiq M. Hossain, and David A. Powell are with School of Engineering and Information Technology, University of New South Wales (UNSW), Canberra, Australia.}
\thanks{Ashif A. Fathnan, Dadin Mahmudin, and Yusuf N. Wijayanto are with Research Center for Electronics and Telecommunication, Indonesian Institute of Sciences (LIPI), Bandung, Indonesia}
\thanks{Email of corresponding author: a.fathnan@student.unsw.edu.au}
\thanks{This work was financially supported by the Indonesia Endowment Fund for Education (LPDP) (PRJ-1081/LPDP.3/2017)}}

\markboth{}
{Fathnan \MakeLowercase{\textit{et al.}}: Achromatic Microwave Focusing Metasurface in Reflection}

\maketitle

\begin{abstract}
We report the characterization of an achromatic focusing metasurface at oblique incident angles. We show that in addition to the inherent off-axis aberrations that occurs due to the hyperbolic phase profile of the metasurface, the focusing performance is significantly degraded due to the meta-atoms' angular dispersion. To obtain insights into how the angular and spectral bandwidth of meta-atoms relate to the metasurface focusing performance, point-dipole models are used which incorporate different aspect's of the meta-atoms' angular response. It is emphasized that despite the meta-atoms being designed under the assumption that they support a single dipolar resonance, other resonances exist within the meta-atom geometry and become stronger at oblique incidence. These resonances disturb the designed phase and amplitude responses, resulting in lower focusing efficiency at higher incident angles. The modelling of higher order modes leads to good agreement with the experimental measurements, confirming that angular dispersion of the meta-atoms is the dominant mechanism in determining off-axis aberrations.
\end{abstract}


\IEEEpeerreviewmaketitle

\section{Introduction}

\begin{figure}[t]
    \centering
    \includegraphics[width=\linewidth]{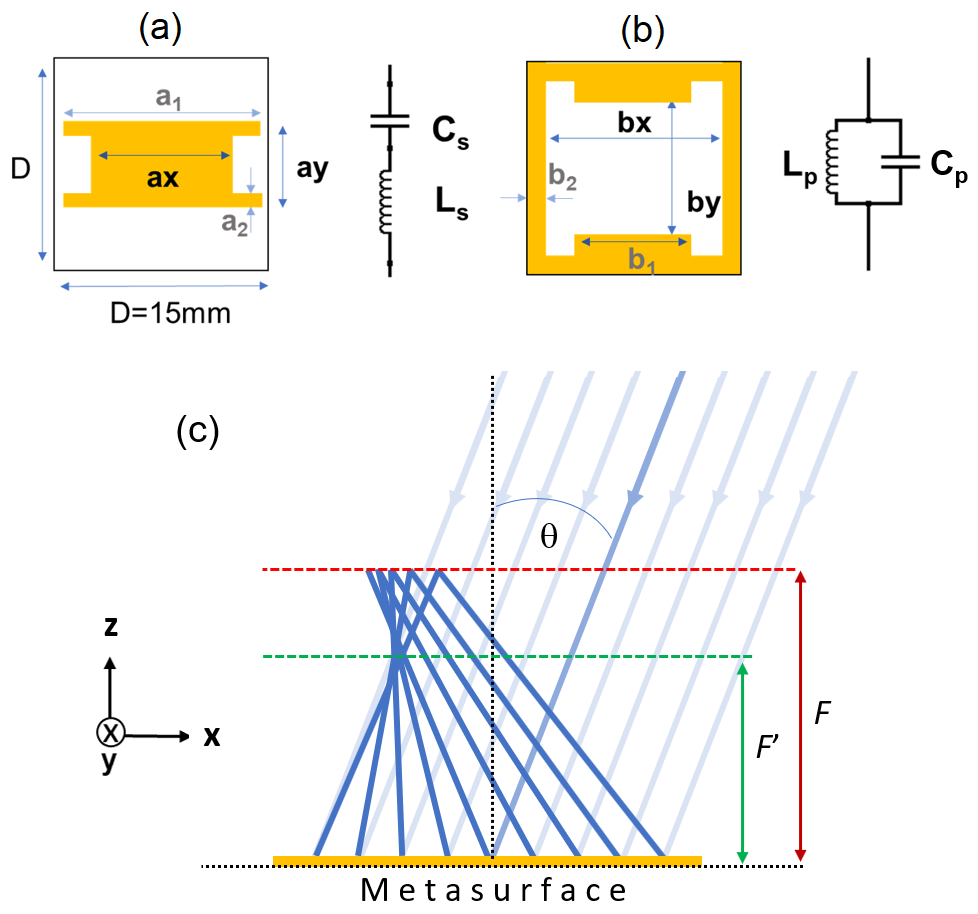}
    \caption{Meta-atom configuration; (a) A dog-bone structure equivalent to a series LC circuit, (b) an inverse dog-bone structure, equivalent to  a parallel LC circuit. (c) Illustration of a reflective focusing metasurface at an oblique incidence. $F$ is the focal spot at a normal incidence and $F'=F\cos\theta$ is the titled focal spot.}
    \label{fig:meta_atom_offaxis}
\end{figure}

\IEEEPARstart{M}{etasurfaces} have attracted high interest due to their ability to provide arbitrary electromagnetic wavefront control. Among many reported works, metasurface synthesis based on judicious engineering of surface impedances has shown to be an effective and rigorous method, enabling realization of flat lenses \cite{fathnan2020achromatic}, prisms \cite{tsilipakos2020squeezing}, axicons \cite{li2014diffraction} and various non-conventional composite surfaces \cite{hadad2015space,he2018waveguide,chen2019design}.
In microwave and millimeter-wave regimes, focusing metasurfaces which maintain high performance over a wide range of incident angles have potential applications, including wireless power transfer \cite{yu2018design,smith2017analysis}, where they are designed to converge plane-waves from various incident angles. Wide angle focusing metasurfaces are also needed as beam steering devices used in automotive radars \cite{schoenlinner2002wide,saleem2017lens}, where directive beams are obtained by changing the position of the feeding source. Despite the need to operate at oblique incidence, the performance of metasurfaces typically degrades compared to normal incidence, which could become a major challenge, hindering their practical applications.

When operating at oblique incidence, inhomogeneous metasurfaces such as metalenses and focusing mirrors suffer degradation due to inherent monochromatic aberrations. The monochromatic aberrations have been well-studied in conventional lenses \cite{wyant1992basic} and diffractive optics \cite{young1972zone,delano1974primary}, where it is known that the hyperbolic phase distribution only results in aberration-free focusing at normal incidence. Therefore, for any focusing devices based on a hyperbolic phase profile, there is inherent degradation at oblique incidence. This is illustrated in Fig.~\ref{fig:meta_atom_offaxis}(c) where a plane-wave impinges upon a metasurface mirror with incident angle $\theta$ causing the focal spot to be off-axis with reduced accumulated power. The degradation effects observed in the focal plane $z=F$ are called Seidel aberrations, and the most important components are coma, astigmatism and curvature \cite{wyant1992basic}. Recently, it has been highlighted that the imaging performance of single-element dielectric metalenses is severely degraded due to coma \cite{liang2019high,decker2019imaging}. Efforts to reduce these inherent off-axis aberrations of focusing metasurfaces have been proposed, such as by modifying the phase profile with additional correcting polynomials, however this introduces another undesired effect of spherical aberrations which further reduces the focusing power at the focal spot \cite{kalvach2016aberration}. Other efforts are by using aplanatic structures \cite{aieta2013aberrations}, and combining two metasurface lenses to create a doublet \cite{arbabi2016miniature}. However, these works add significant complexity and resulted in bulky structures. Additionally, these works only addressed the inherent monochromatic abberations, where in realization, \textit{the meta-atom angular dispersion may significantly contribute to the focusing degradation}. Previous studies have observed strong angular dispersion on thin resonant meta-atoms \cite{minovich2010tilted,albooyeh2014resonant,qiu2018angular,zhang2020controlling} where oblique incident illumination influences the amplitude and phase responses.

The angular dispersion of meta-atoms has been characterized in various works, including Ref.~\cite{minovich2010tilted}, where a fishnet metamaterial structure is analyzed. The results show that the metasurface structure is more sensitive to oblique incidence when the electric field is normal to the metal plates, than when it is parallel to them. In Ref.~\cite{albooyeh2014resonant}, a study on the oblique incidence of resonant metasurfaces was conducted considering the lattice symmetry of the meta-atoms. It was shown that the meta-atoms which consist of stacked square-patches have mainly two resonance mode, i.e. a dipole mode and a quadrupole mode. Both resonance modes behave differently in the presence of oblique incidence, depending on the lattice symmetry being implemented. For the amorphous lattice metasurface, the dipole resonance is strongly affected by the oblique incidence, while the quadrupole resonance remains unchanged. Additional work on characterizing oblique incidence in metasurfaces can be found in Refs.~\cite{yazdi2017analysis,qiu2018angular}, where theoretical models are developed to predict angular dispersion profile of the meta-atoms. Despite being able to explain the cause of amplitude and phase changes at oblique incidence, the works reported in \cite{minovich2010tilted,albooyeh2014resonant,qiu2018angular} only considered homogeneous metasurfaces, and did not evaluate the performance impact on focusing inhomogeneous metasurfaces.

In this paper we characterize the performance of a broadband focusing metasurface at oblique incidence, showing the contribution of both inherent monochromatic aberrations and meta-atom angular dispersion. We use a broadband metasurface based on the design procedure presented in Ref.~\cite{fathnan2020bandwidth}, utilizing dog-bone and inverse dog-bone structures, in a metallic-dielectric meta-atom architecture. Experimental verification is performed in the X-band microwave frequency range, considering an incident plane wave with transverse electric (TE) polarization. To understand the contributions of both meta-atom angular dispersion and inherent off-axis aberrations, numerical and analytical calculations are conducted using a point dipole model. Our study reveals that in addition to the designed dipole resonances, the meta-atoms have higher order modes, with opposite tendency of frequency shift at oblique TE-incidence. This leads to non-negligible effects of resonance coupling at higher incident angles which significantly degrades the efficiency of a broadband focusing metasurface. Furthermore, this study gives insights into how the angular dispersion of metasurfaces can be reduced, by suppressing this coupling between multipolar modes of the constituting meta-atoms.

\begin{figure}[t]
    \includegraphics[width=\linewidth]{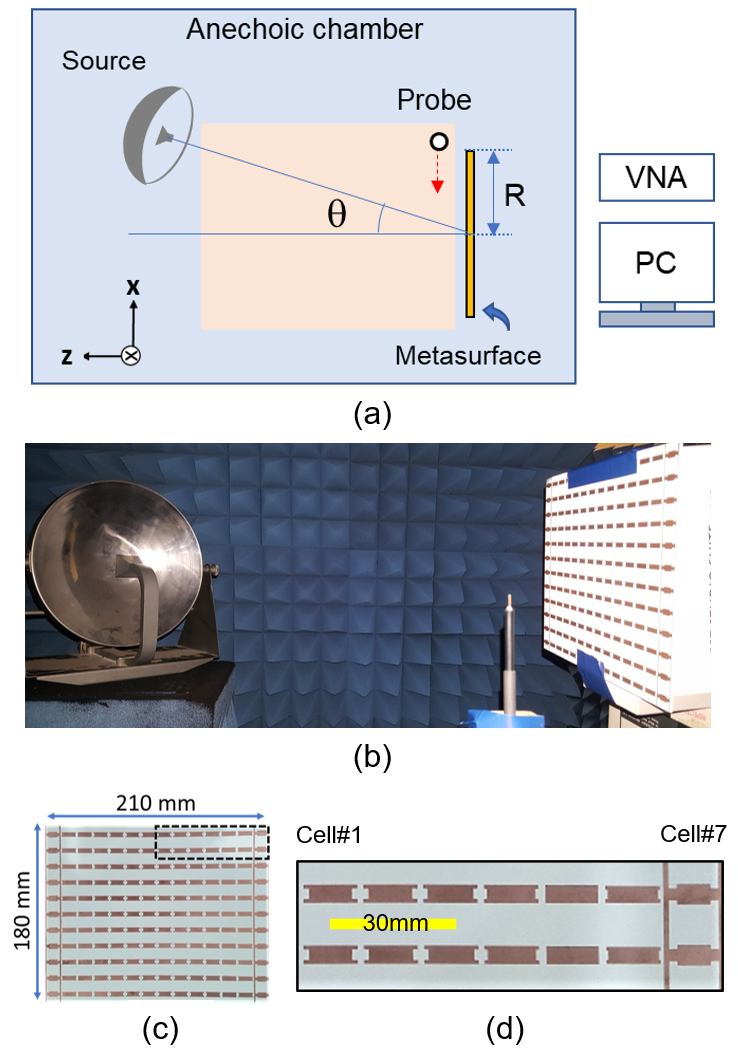}
    \caption{(a) Schematic and (b) photograph of the near-field measurement setup inside an anechoic chamber. (c) Photograph of the metasurface sample with size annotated. (d) Close up view of seven meta-atoms making one radius of the metasurface focusing mirror. Yellow line indicates 30\,mm equals to the wavelength at the center frequency.}
    \label{fig:sample_meassetup}
\end{figure}

\section{Design, Simulation and Measurement Results \label{sec:meas_sim}}

To understand the effect of oblique incidence, we synthesize and fabricate a broadband metasurface using the method reported in Ref.~\cite{fathnan2020bandwidth}. The metasurface is realized by tailoring the Q-factor of the resonant structures, which is equivalent to modifying the parameters of series and parallel LC circuits. Each meta-atom's equivalent L and C values are obtained by fitting the surface impedance and its first order frequency derivative at the center frequency. The design center frequency is 10\,GHz or center wavelength of $\lambda_0$=30\,mm, with the amplitude focal length $F$=190\,mm (or the phase focal length $F_p$=300\,mm) and the aperture radius R=105\,mm. The metasurface is designed to have achromatic focusing performance with 12\% fractional bandwidth, defined here as the bandwidth over which phase deviation from the design is less than 0.2$\pi$ (see Fig.~\ref{fig:ucphase} of the Appendix). 
Within one radius, the metasurface is discretized into 7 unit cells, consisting of six dog-bone structures (series-LC circuits) as shown in Fig.~\ref{fig:meta_atom_offaxis}(a), and one inverse dog-bone structure (a parallel-LC circuit) as shown in Fig.~\ref{fig:meta_atom_offaxis}(b). These metallic structures are 35\,$\mu$m copper traces etched on top of Rogers RO4360G2 dielectric substrate with thickness of 1.52\,mm, permittivity of 6.15, and dissipation factor of 0.0038. The dielectric substrate is backed by a full plane copper layer, yielding reflective metasurface operation. Details of the meta-atom parameters are provided in Table~\ref{tab:meta-atom-para} of the Appendix.

The fabricated sample of the focusing metasurface is shown in Fig.~\ref{fig:sample_meassetup}(c) with the close-up view of one metasurface radius where the cell numbers are indicated. A near-field measurement setup located inside an anechoic chamber is used to characterize the fields scattered from the metasurface. As shown in Fig.~\ref{fig:sample_meassetup}(a), the metasurface is illuminated with a horn antenna that feeds  a parabolic mirror, to approximate a plane-wave source. A photograph of the measurement setup is shown in Fig.~\ref{fig:sample_meassetup}(b). The scattered fields in front of the metasurface are captured by a moving probe connected to a VNA and controlled by a personal computer. To vary the incident angle, the horn antenna is shifted to different discrete angles $\theta$. To obtain the field reflected from the metasurface, we perform a separate measurement of the incident plane wave in the absence of the sample, and subtract it from the total field.

\begin{figure}[t]
    \includegraphics[width=\linewidth]{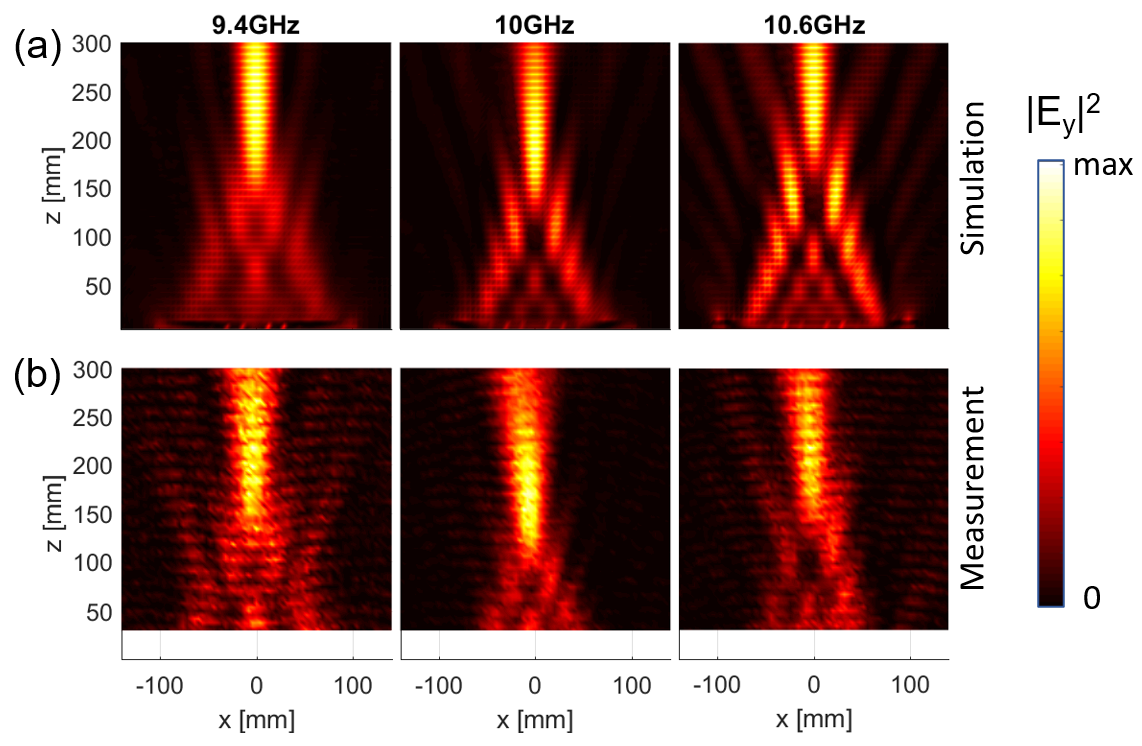}
    \includegraphics[width=\linewidth]{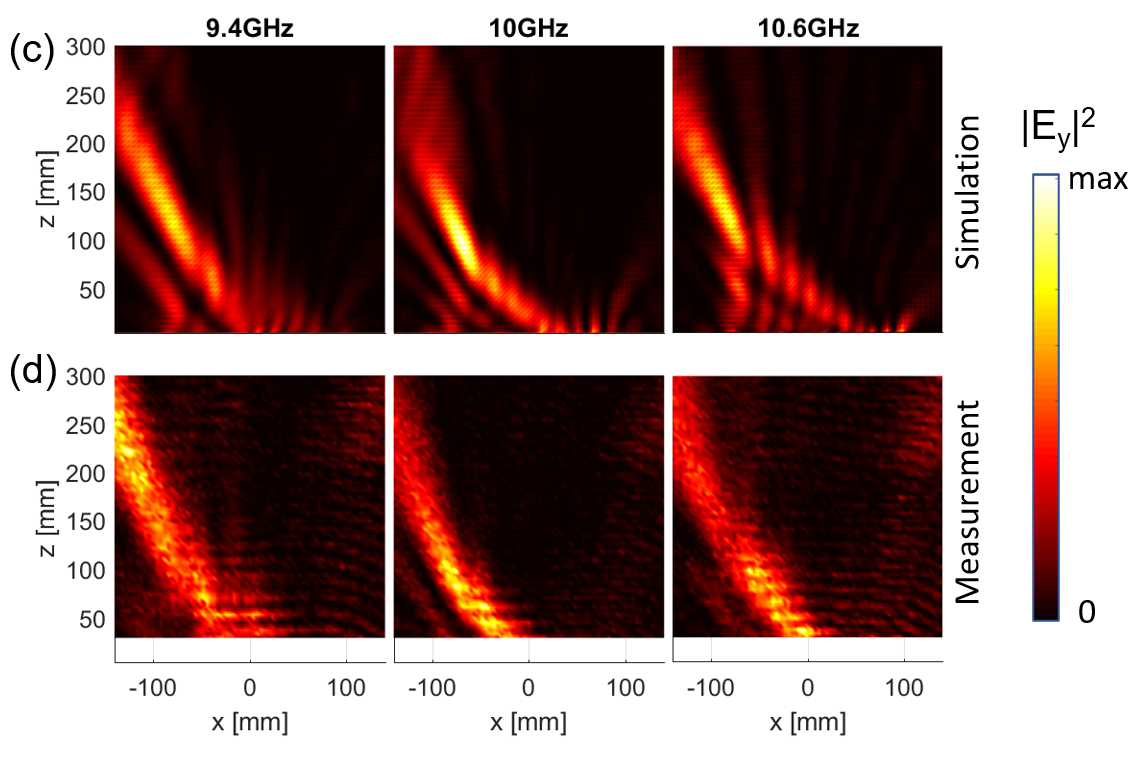}
    \includegraphics[width=\linewidth]{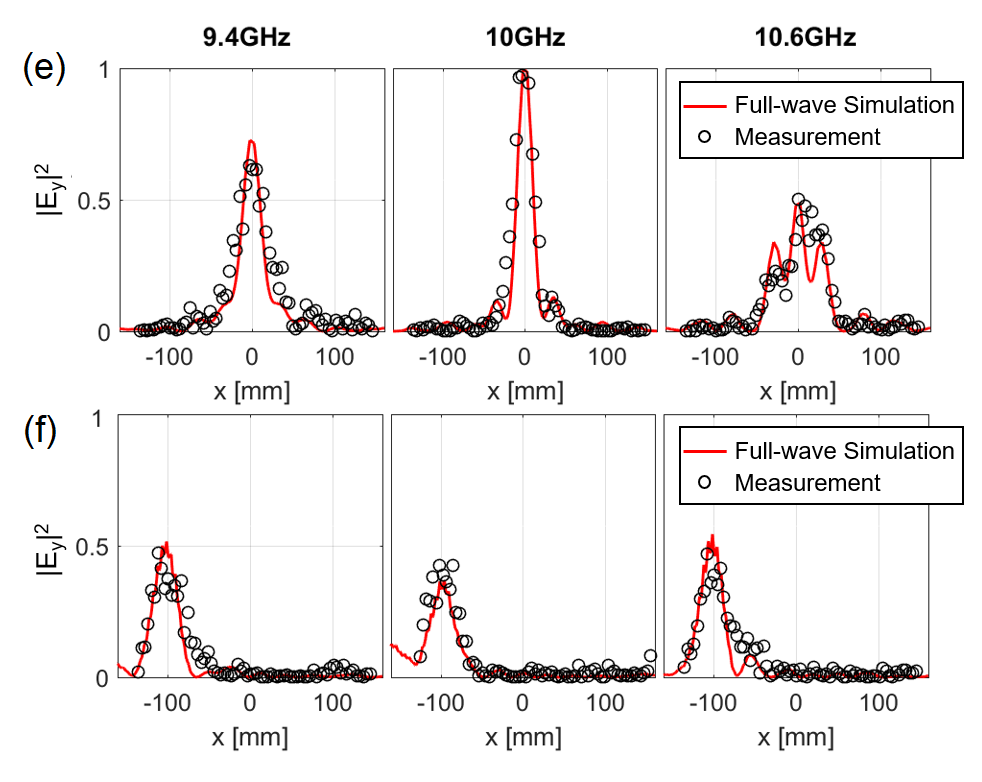}
    \caption{Scattered field intensity from full-wave simulation and measurement of the metasurface within $xz$-plane at three working frequencies for (a)-(b) normal incidence and (c)-(d) oblique incidence with $\theta=30^\circ$. Scattered field intensity along $x$-axis at (e) the focal plane $F$ and (f) at the shifted focal plane $F'$ with $\theta=30^\circ$ comparing the result from measurement (circles) and full-wave simulation (red-lines).}
    \label{fig:measurement_vsfullwave}
\end{figure}

In Fig.~\ref{fig:measurement_vsfullwave} the scattered field intensity ($|E_y|^2$) from (a) CST full-wave simulation and (b) measurement are compared for a normally incident wave. In Fig.~\ref{fig:measurement_vsfullwave}(c) and (d), the CST full-wave simulation and the measurement results are shown for oblique incidence with $\theta = 30^\circ$. Similar to the measurement, the CST full-wave simulation results presented in Fig.~\ref{fig:measurement_vsfullwave}(a) and (c) show the fields reflected from the metasurface, where the incident plane waves have been subtracted. Since the metasurface is uniform along the $y$-axis, in the simulation, a vertical homogeneous approximation is used by setting electric boundary conditions at the vertical edges of a one-dimension metasurface. The scanning area of the metasurface is between -144\,mm to 144\,mm in the $x$-axis and to a maximum of 300\,mm in the $z$-axis, with the center of the metasurface located at $z=0$, $x=0$. The probe minimum step movement is 4\,mm in both $x$ and $z$ coordinates. Note that since the probe cannot be too close to the metasurface, the measurement starts at $z=32$\,mm.

In order to characterize the focusing properties of the metasurface, we evaluate the field distribution at the focal plane. This method is similar to performing a translational scan at the focal plane, as reported in several works on focusing metasurfaces \cite{zhang2016high,arbabi2016miniature}.
At normal incidence $\theta=0^\circ$, the designed focal plane is located at distance $F$ along the $z$-axis from the metasurface. At oblique incidence, the reflected beam is tilted, resulting in a shifted focal plane $F'=F\cos{\theta}$ where the distance becomes closer to the metasurface, as indicated in Fig.~\ref{fig:meta_atom_offaxis}. The field distribution at the shifted focal plane $F'$ is used 
since the maximum field intensity is expected to be at this point according to ray optics arguments. The ideal field distribution from the point dipole calculation can be consistently used as a reference when characterizing off-axis aberrations, as will be shown in Section~\ref{sec:numerical_study}. 
This characterization is relevant for applications where a static local wave source is used, such as in metasurface for far-field beam forming \cite{olk2020huygens,akram2020bi} and reflectarray antennas \cite{abbasi2019millimeter,kiris2019reflectarray} as well as for wireless power transfer applications \cite{yu2018design,smith2017analysis}. 

Profiles of field intensity at the focal plane are plotted in Fig.~\ref{fig:measurement_vsfullwave} for (e) $\theta=0^\circ$ and (f) $\theta=30^\circ$. Three frequencies are shown, corresponding to the operating bandwidth of the achromatic metasurface between 9.4-10.6\,GHz. All field intensities are normalized to the maximum value when the impinging plane wave is at normal incidence $\theta=0^\circ$ and at the center operating frequency $f_0=10$\,GHz. Good agreement between the measurement and CST full-wave simulation can be observed, where we see that the field intensity degrades as the incident angle changes from $\theta=0^\circ$ to $\theta=30^\circ$. For oblique incidence, the peak is shifted towards off-axis coordinate of around $x=-100$\,mm, consistent with ray optics calculation. We can also see that at normal incidence, the field intensity degrades as the operating frequency shifts further from the center. This indicates that the metasurface has both angular and frequency dependent focusing degradation. To fully characterize the focusing performance over the spectral and angular bandwidth, calculation of the focusing efficiency based on full-width half magnitude (FWHM) of the field intensity profile \cite{arbabi2016miniature} will be shown in Section~\ref{sec:degradation_focusing}.

\section{Numerical Study of the Metasurface Focusing Performance \label{sec:numerical_study}}
\subsection{Non-Dispersive Model \label{sec:field_distb}}
Fig.~\ref{fig:measurement_vsfullwave} shows experimentally and numerically that when a metasurface is excited at oblique incidence, the focusing performance degrades. Here we investigate the contribution of this degradation from both the inherent monochromatic aberrations and angular dispersion of the meta-atom. To understand the effect of inherent monochromatic aberrations, we conduct a numerical study to replicate the field distribution of the metasurface using a point dipole model. By approximating each meta-atom as a point dipole, the field distribution from the metasurface can be numerically reconstructed \cite{li1981focal,yang2014efficient,born1999principles}. Considering the metasurface as a one dimensional array of dipoles arranged along the $x$-axis, as shown in Fig.~\ref{fig:meta_atom_offaxis}, at normal incidence, the field scattered by the metasurface can be formulated as
\begin{equation}
E_y{(x,z)}=\sum_{x,z}\frac{A(x_m,\omega)}{\sqrt{r}}e^{-j(k_0r+\Phi_r(x_m,\omega))}. \label{eq:pointdipole_normal}
\end{equation}
Here, $A$ is the meta-atom amplitude response and $\Phi_r$ is the phase response, $r$ is the distance from the meta-atom location to arbitrary point $x, z$ within the $xz$-plane in front of the metasurface and $k_0=2\pi/\lambda$ is the wave number. Both the amplitude $A$ and the phase response $\Phi_r$ of the meta-atoms depend on frequency $\omega$ and on location $x_m$, which is the distance of the meta-atom relative to the center of the metasurface. 
Under the assumption that angular dispersion of the meta-atom is negligible, the amplitude $A$ is always one and the phase $\Phi_r$ has a hyperbolic profile following 
\begin{equation}
\Phi_{r}(x_m,\omega)=k_0 (\sqrt{x_m^2+F_p^2}-F_p)+ \Phi_0(\omega).
\label{eq:hyperbolic_phase}
\end{equation}
$F_p$ is the design phase focal length of the metasurface and $\Phi_0$ is an additional phase which can be added without changing the focusing performance. To account for the effect of intrinsic monochromratic aberrations, Equation \eqref{eq:pointdipole_normal} can be modified to include the different phase and amplitude at which each meta-atom is excited at oblique incidence. The phase changes as $\Phi_\theta=k_0 x_m \sin{\theta}$ due to the varying optical path length. The amplitude changes are due to the change of flux, with a factor of $\sqrt{\cos(\theta})$ added to the original amplitude. Both of these effects can be incorporated into the point dipole model, resulting in
\begin{equation}
E_y{(x,z)}=\sum_{x,z}\frac{A(x_m,\omega)\sqrt{\cos(\theta)}}{\sqrt{r}}e^{-j(k_0r+\Phi_r(x_m,\omega))+\Phi_\theta}. \label{eq:pointdipole_oblique}
\end{equation}
By adding the phase and amplitude changes,  the field distribution predicted by Eq.~\eqref{eq:pointdipole_oblique} accounts for inherent monochromatic aberrations. The wavefront deformation can be described via a Taylor expansion of the phase profile where the inherent monochromatic aberrations of coma, astigmatism and curvature can be analytically distinguised \cite{aieta2013aberrations,ginn2010monochromatic,decker2019imaging}. We denote this formulation as the non-dispersive model, since it neglects both the meta-atom angular and spectral dispersion.  

\begin{figure}
    \centering
    \includegraphics[width=\linewidth]{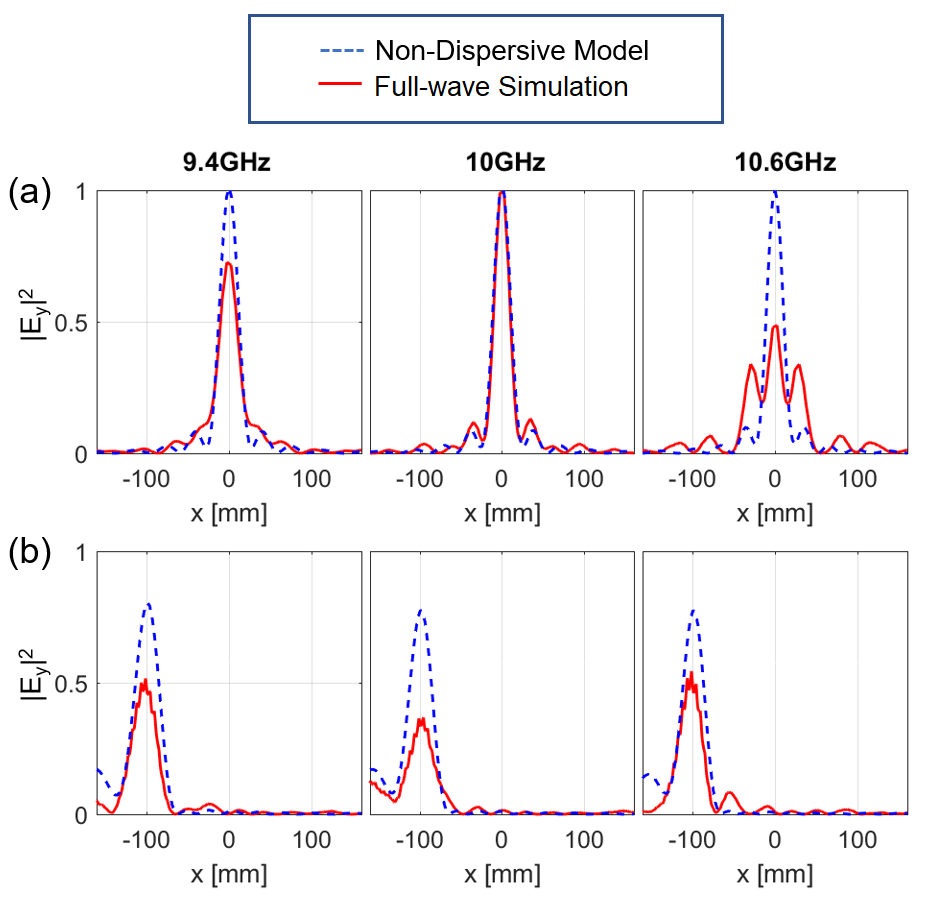}
    \caption{Comparison of field intensity from the point dipole calculation of meta-atoms based on a non-dispersive model (dashed lines) and the field intensity from metasurface full-wave simulation result (red lines) along $x$-axis at (a) the focal plane $F$ and (b) at the shifted focal plane $F'$ with $\theta=30^\circ$.}
    \label{fig:fw_vs_nondisp}
\end{figure}

The field distribution calculated by Eq.~\eqref{eq:pointdipole_oblique} at $F$ and $F'$ for several frequencies are plotted in Fig.~\ref{fig:fw_vs_nondisp} and compared with the result from the full-wave numerical simulation. It can be seen from the dashed line of Fig.~\ref{fig:fw_vs_nondisp}(b) that by using this point dipole calculation, intrinsic aberration in the form of coma is clearly observed as a second peak appears around $x=-140$\,mm. However, the calculated field distribution does not agree with the full-wave simulation result where a stronger degradation is observed. Additionally, using this non-dispersive calculation, the field distribution is the same for all frequencies, which deviates from the simulated results.
Therefore, in order to get better agreement, a key method introduced in this numerical study is to incorporate the meta-atom angular dispersion into Eq.~\eqref{eq:pointdipole_oblique}, such that the ideal amplitude and phase responses are replaced by a more realistic quantity having both angular and spectral dependency, i.e.~$A(\omega,x_m,\theta)$ and $\Phi_r(\omega,x_m,\theta)$. In doing so, we consider first, a single-mode dispersion model obtained by LC-circuit analysis (Section~\ref{sec:angdip_lcmodel}), and then a multi-mode dispersion model obtained by full-wave numerical simulation (Section~\ref{sec:angdip_fullwave}).  

\subsection{Single Mode Dispersion Model \label{sec:angdip_lcmodel}}
\begin{figure}[t]
    \includegraphics[width=\linewidth]{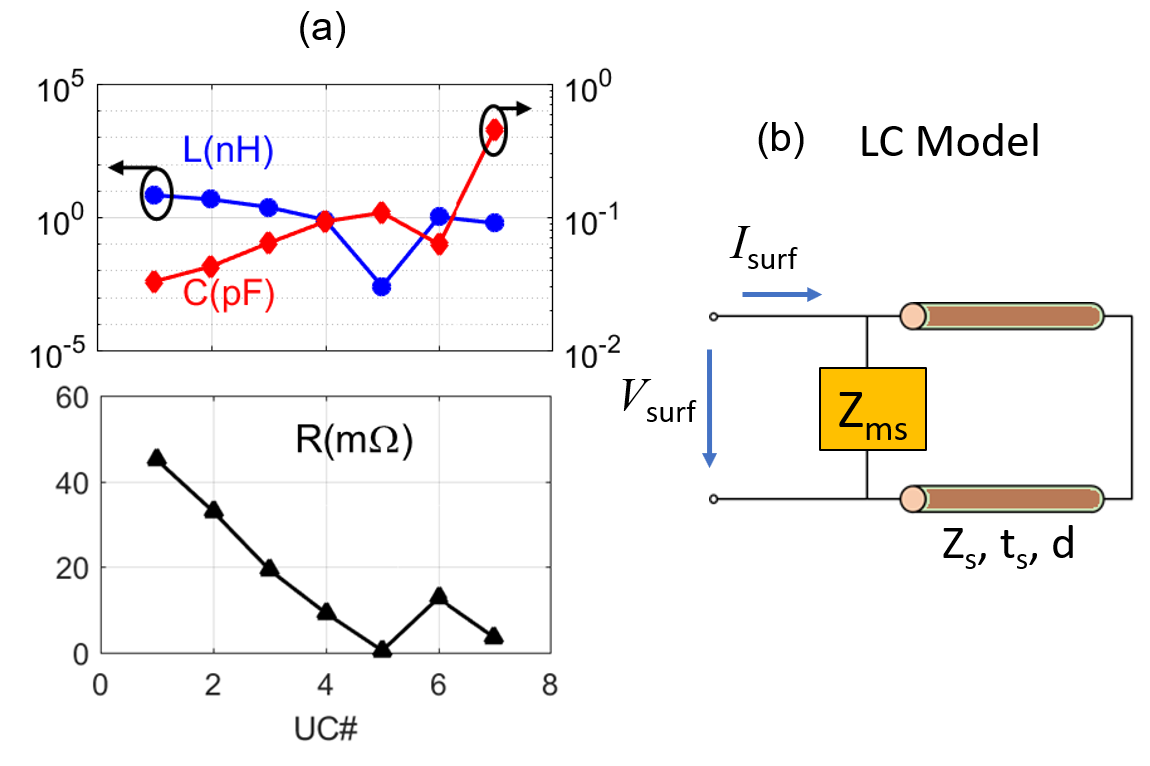}
    \caption{(a) RLC parameters for the designed focusing metasurface based on the LC-model. (b) Equivalent circuit schematic of the meta-atom, where meta-atom impedance ($Z_\mathrm{ms}$) is connected to a grounded transmission line.}
    \label{fig:LC_model}
\end{figure}

In obtaining the angular dependence of the amplitude and phase responses, first we consider analytical calculation from an ideal isolated LC resonance. 
In a thin metallic-dielectric meta-atom structure, the relation between tangential fields can be described in terms of surface impedances. An equivalent circuit can be used to analyze the overall field interaction within each meta-atom, where as shown in Fig.~\ref{fig:meta_atom_offaxis}(a)-(b), the meta-atoms here are equivalent to either a series or parallel LC circuit. 
By considering the metallic losses as an additional resistance connected in series to the inductance, equivalent impedance of the meta-atom $Z_\mathrm{ms}$ can be  formulated as, 
\begin{align}
Z_\mathrm{ms}&=\left(\frac{1}{j\omega L_p + R} +j\omega C_p\right)^{-1}, &\mathrm{(LC\ Parallel)} \label{eq:y_lc_parallel} \\
Z_\mathrm{ms}&=j\omega L_s +R+ \frac{1}{j\omega C_s}, &\mathrm{(LC\ Series)} \label{eq:y_lc_series}
\end{align}
Here, the resistance $R$ is obtained by fitting to the unit-cell numerical simulation results, while $L$ and $C$ are the inductance and capacitance obtained from the design procedure outlined in Ref.~\cite{fathnan2020bandwidth}. For the focusing metasurface as realized in Section~\ref{sec:meas_sim}, the L, C and R profiles are plotted in Fig.~\ref{fig:LC_model}(a). Based on this LC-model, as illustrated in Fig.~\ref{fig:LC_model}(b) the meta-atom impedance $Z_\mathrm{ms}$ 
appears as a shunt load on a transmission line, which represents propagation through the dielectric substrate of thickness $d$. The other port of the transmission line is grounded, representing the metallic ground plane. By using an ABCD transfer matrix to model the transmission line, we can relate the surface voltages and currents to those on the ground plane as,

\begin{equation}
\begin{bmatrix}
V_\mathrm{surf}\\
I_\mathrm{surf}
\end{bmatrix}
=
\begin{bmatrix}
1 & 0\\
\mathrm{Y_\mathrm{se}} & 1
\end{bmatrix}
\begin{bmatrix}
A & B\\
C & D
\end{bmatrix}
\begin{bmatrix}
0\\
I_\mathrm{gnd}
\end{bmatrix}. \label{eq:abcd_matrix_ma}
\end{equation}
Here, $A=D=\cos\left(\omega t_s\right)$, $B=jZ_s\sin\left(\omega t_s\right)$, $C=j\frac{1}{Z_s}\sin\left(\omega t_s\right)$ are the transmission line parameters. Propagation through the substrate is expressed as a time delay $t_s=d\sqrt{\epsilon_s}/c$ and the substrate impedance is $Z_\mathrm{s}=\sqrt{\frac{\mu_0}{\epsilon_0\epsilon_s}}=\frac{\eta}{\sqrt{\epsilon_s}}$, where $\epsilon_0$ and $\mu_0$ are free space permittivity and permeability and $c$ is the speed of light in vacuum. The substrate thickness $d$, permeability $\mu_s$ and permittivity $\epsilon_s$ are specified, based on the dielectric material properties (Roger RO4360G2) given in Section~\ref{sec:meas_sim}. At oblique incidence, the wave propagation within the substrate changes, where as a result, the substrate impedance increases and the propagation delay decreases. Considering a transverse electric (TE) propagating wave, at oblique incidence the substrate impedance and the substrate time delay are
\begin{equation}
\begin{aligned}
Z_\mathrm{s}(\theta)=\frac{\eta}{\sqrt{\epsilon_s}\cos{\theta}} \\
t_s(\theta)=\frac{d \sqrt{\epsilon_s} \cos{\theta}}{c} \label{eq:sub_para_oblique}
\end{aligned}
\end{equation}
Using the above angle dependent substrate parameters, we obtain the angle dependent meta-atom reflection coefficient $S_{11}$. By substituting Eq.~\eqref{eq:sub_para_oblique} into Eq.~\eqref{eq:abcd_matrix_ma}, and defining the input impedance $Z_{11}$ as the ratio between the equivalent surface voltage $V_\mathrm{surf}$ and current $I_\mathrm{surf}$, we obtain,
\begin{figure}[t]
    \centering
    \includegraphics[width=\linewidth]{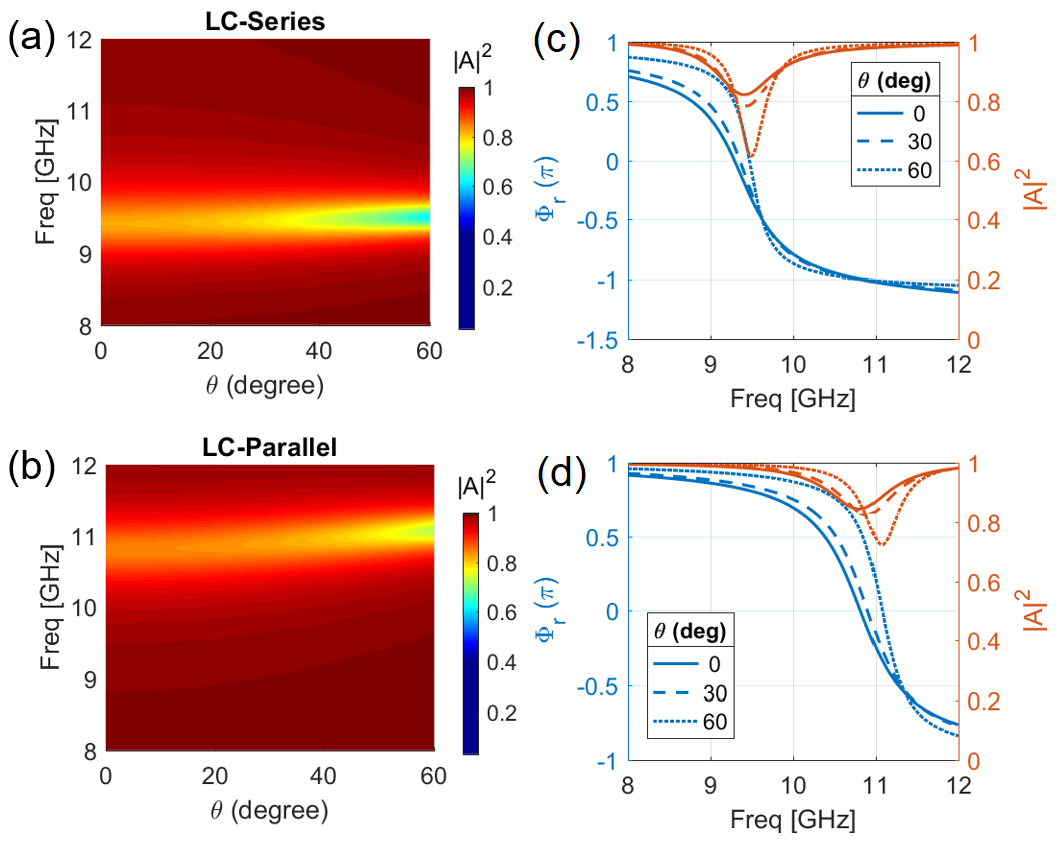}
    \caption{ The angular disperion of the meta-atom reflection intensity based on an LC-model for; (a) and (c) a dog-bone structure used as cell \#1, (b) and (d) an inverse dog-bone structure used as cell \#7.}
    \label{fig:angular_disp_LCmodel}
\end{figure}
\begin{equation}
Z_{11}(\omega,\theta)=\frac{jZ_s\sin(\omega t_s)}{jY_{se}Z_s\sin(\omega t_s)+\cos(\omega t_s)} \\
\label{eq:z11_ts_zs}
\end{equation}
This is related to the reflection coefficient $S_{11}$ as
\begin{equation}
S_{11}(\omega,\theta)=\frac{Z_{11}(\omega,\theta)-\eta/\cos{\theta}}{Z_{11}(\omega,\theta)+\eta/\cos{\theta}} \\
\label{eq:s11_ts_zs}
\end{equation}
This reflection coefficient includes both angular and spectral dispersion of the substrate, hence it can be incorporated into Eq.~\eqref{eq:pointdipole_oblique} to obtain a realistic field distribution from the metasurface using $A(\omega,\theta)=|S_{11}|$ and $\Phi_r(\omega,\theta)=\angle S_{11}$. Note that L and C here \textit{do not change} with the incident angles since they are obtained from an ideal LC-circuit model. This formulation has an implicit assumption that the structure supports only a single dipolar mode, which can be represented by the equivalent circuit model. An example of the meta-atom angular and spectral dispersion based on this model is shown in Fig.~\ref{fig:angular_disp_LCmodel} for (a) dog-bone structure of cell \#1, and (b) inverse dog-bone structure of cell \#7. It is clear that in both cases the meta-atom resonances shift to higher frequencies as the incident angle increases from 0$^\circ$ to 60$^\circ$. Fig.~\ref{fig:angular_disp_LCmodel}(c) and (d) show amplitude and phase spectra for three incident angles, emphasizing the more extreme variation of amplitude and phase at higher angles.

\begin{figure}[t]
    \centering
    \includegraphics[width=\linewidth]{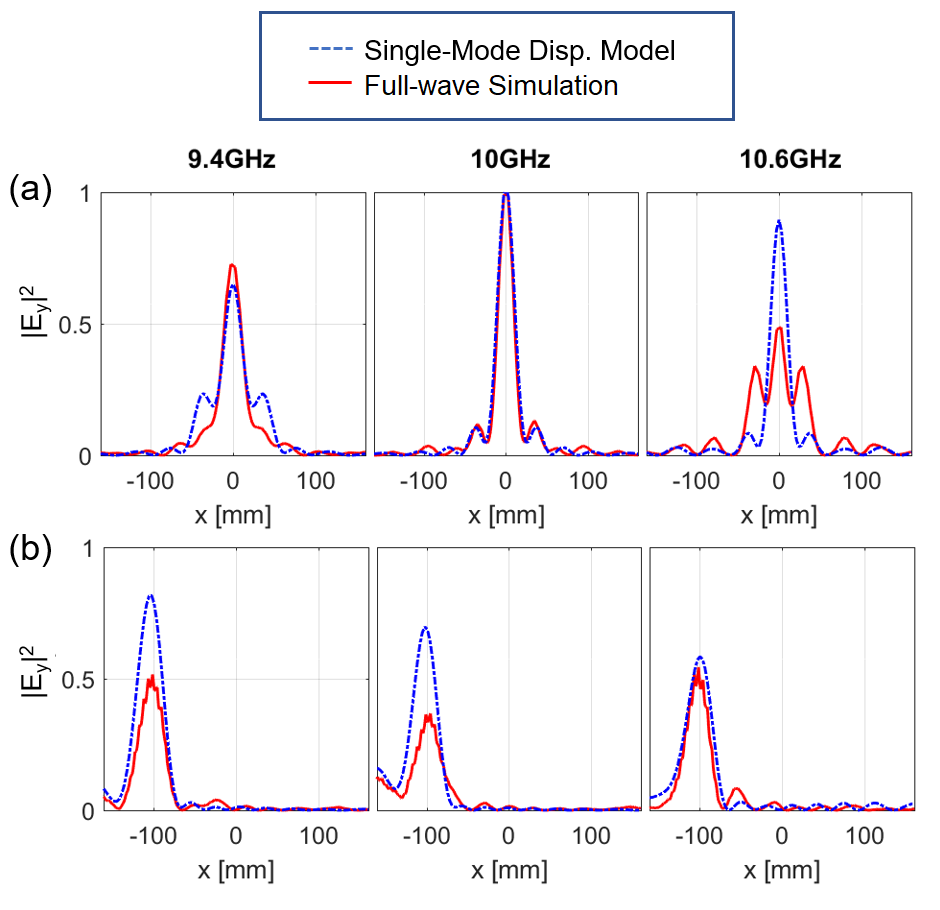}
    \caption{Comparison of field intensity from the point dipole calculation of meta-atoms based on a single-mode dispersion model (dashed lines) and the field intensity from metasurface full-wave simulation result (red lines) along $x$-axis at (a) the focal plane $F$ and (b) at the shifted focal plane $F'$ with $\theta=30^\circ$.}
    \label{fig:fw_vs_singlemode}
\end{figure}

\begin{figure}
    \centering
    \includegraphics[width=\linewidth]{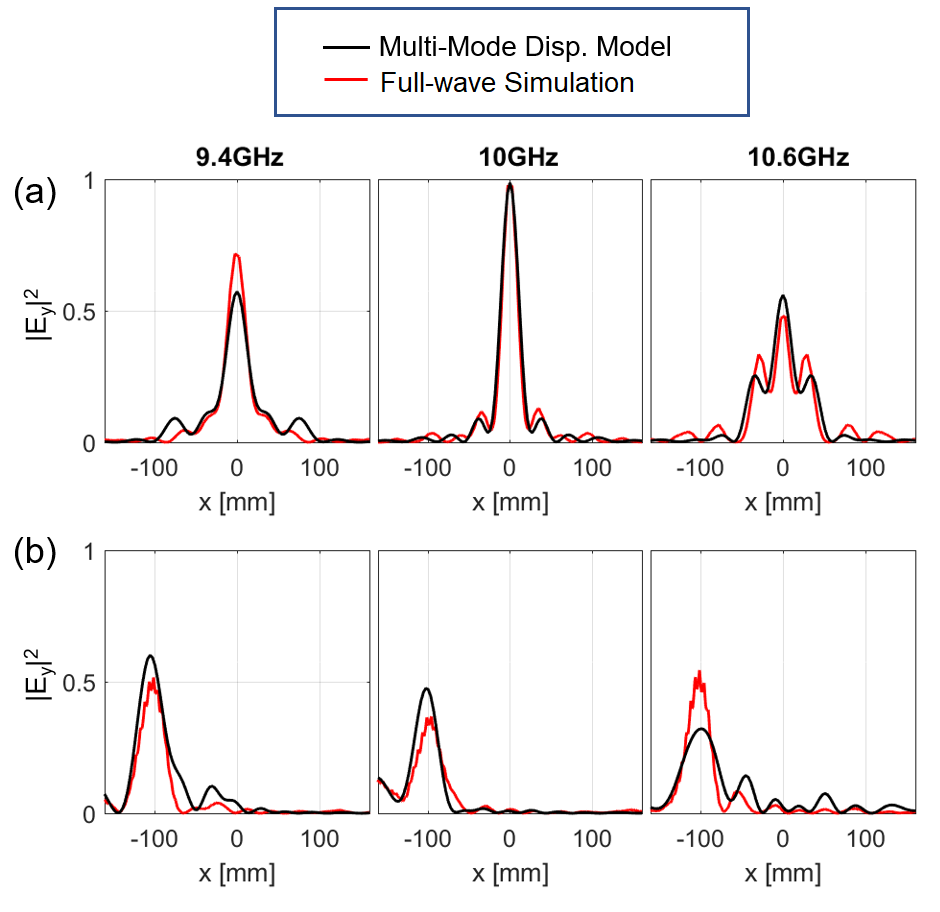}
    \caption{Comparison of field intensity from the point dipole calculation of meta-atoms based on a multi-mode dispersion model (black lines) and the field intensity from metasurface full-wave simulation result (red lines) along $x$-axis at (a) the focal plane $F$ and (b) at the shifted focal plane $F'$ with $\theta=30^\circ$.}
    \label{fig:fw_vs_multimode}
\end{figure}

Using this single dipolar mode assumption, the field profiles at $F$ and $F'$ for several frequencies are plotted in Fig.~\ref{fig:fw_vs_singlemode} and compared with the result from the full-wave numerical simulation. It can be seen that even at the normal incidence, the field profile degrades when the frequency is further from the design (see dashed lines of Fig.~\ref{fig:fw_vs_singlemode}(a)), indicating the spectral dispersion of the LC-resonance model. However, the calculated field distribution still does not agree well with the simulation result. This is due to the LC-resonance model assuming that only a single dipolar mode is excited, hence it does not capture all the physics of the meta-atom. As described in previous works \cite{albooyeh2014resonant,liu2017high,liu2010planar}, resonant meta-atoms may have higher order modes excited at oblique incidence, which introduce further deviation to the amplitude and phase responses and contribute to the focusing degradation.

\subsection{Multi-Mode Dispersion Model \label{sec:angdip_fullwave}}
\begin{figure*}[t]
    \centering
    \includegraphics[width=\linewidth]{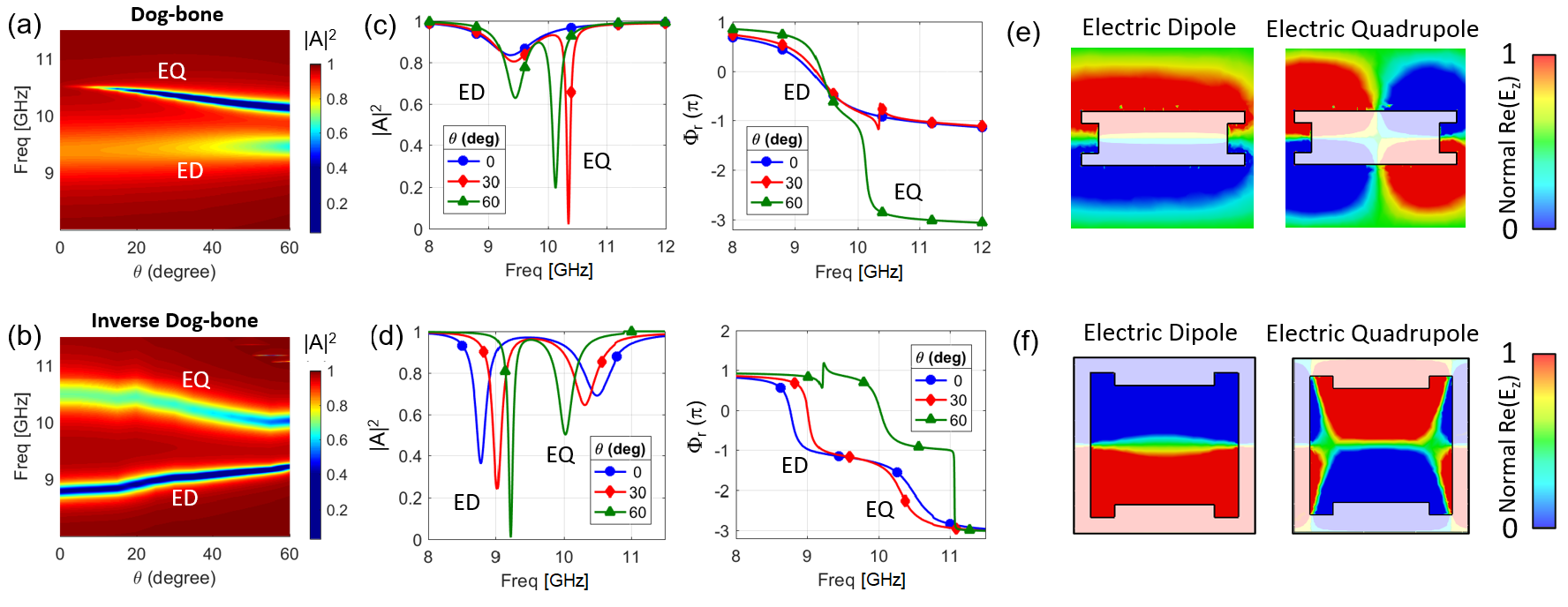}
    \caption{ The angular disperion of the meta-atom reflection intensity for (a) a dog-bone structure used as cell \#1 and (b) an inverse dog-bone structure used as cell \#7, where contribution from electric dipole (ED) and electric quadrupole (EQ) modes are indicated. (c) and (d) show the reflection intensity and phase profile for chosen incident angles, for the dog-bone and inverse dog-bone structure respectively. The simulated dipole and quadrupole mode of normal electric fields shown for (e) the dog-bone structure and (f) the inverse dog-bone structure}
    \label{fig:angular_disp_CST}
\end{figure*}

Due to the disagreement of the single-mode dispersion model with the simulation results, here, the meta-atom is modeled by using unit-cell CST simulation. The key difference introduced in this approach is that a full-wave solution is used to find the angular dispersion of each meta-atom. This full-wave solution provides more accurate results, where the amplitude and phase responses are obtained by considering the realistic meta-atom geometrical parameters and losses of both the dielectric and metallic materials. Exemplary results are shown in Fig.~\ref{fig:angular_disp_CST} for (a) dog-bone structure of cell \#1, and (b) inverse dog-bone structure of cell \#7. It is clear that instead of the single resonance predicted by the equivalent circuit model, both meta-atoms support multi-mode dispersion with two resonances found near or within the operating bandwidth. These resonances shift frequency in opposite directions as the incident angle increases from 0$^\circ$ to 60$^\circ$. In the dog-bone structure, the second resonance is clearly observable only after $\theta$=15$^\circ$, while in the inverse dog-bone structure it appears even at normal incidence.  Figure \ref{fig:angular_disp_CST}(c) and (d) show the amplitude and phase spectra for three incident angles, where we see that the two resonances in each meta-atom become closer as the incident angle increases. This causes severe change of the phase and amplitude responses, which degrades the focusing performance at higher incident angles.

The focusing performance is shown in Fig.~\ref{fig:fw_vs_multimode}, where the field distribution at $F$ and $F'$ are plotted. This field distribution is obtained by substituting the simulated multi-mode dispersion of the meta-atoms into Eq.~\eqref{eq:pointdipole_oblique}. The result shows that the point dipole model (black lines) agrees well with the simulation of the metasurface (red lines). A small level disagreement can still be observed, which is expected to exist, given the inhomogeneous and finite properties of the realized metasurface structures, while the point dipole model is based on a unit-cell meta-atom simulation with infinite homogeneous approximation. Despite this small discrepancy, the predicted field distribution shows much better agreement than using only single dipolar model from the meta-atom equivalent LC circuit. This result demonstrates that the meta-atom angular dispersion dominates the degradation of the metasurface focusing performance, with the undesired resonances being the most significant cause.

To understand the nature of these resonances, in Fig.~\ref{fig:angular_disp_CST}(e) and (f) we plot the normal component of the electric fields from the two observed resonances at the plane of the meta-atom. 
It can be identified that the first resonance of the dog-bone structure is a dipole mode and the second resonance is an electric quadrupole mode (see Fig.~\ref{fig:angular_disp_CST}(e)). The electric quadrupole mode exists due to the alternating charges supported by the meta-atom which can be excited only at oblique incidence. Similarly, for the inverse dog-bone structure, the electric field profiles in Fig.~\ref{fig:angular_disp_CST}(f) show that the first resonance is an electric dipole mode and the second resonance is an asymmetric electric quadrupole mode. The second resonance here appears even at the normal incidence, indicating that it is not a purely quadrupolar mode, but is a hybrid resonance with a dipolar component. 
Unlike the purely dipole resonance that appears around 9\,GHz, the hybrid electric quadrupole resonance can be controlled by manipulating the geometry of the capacitive patch ($b_y$ and $b_1$ in Fig.~\ref{fig:meta_atom_offaxis}(b)), and has an advantage of low Q-factor suitable for broadband operation. This second low Q resonance is also observed in other similar structures, for example electric split-rings above a ground plane \cite{tao2008highly,tao2010dual}, non-uniform square patches \cite{zhang2013multispectral}, and double crosses structures \cite{he2011dual}.

\section{Degradation of Focusing Performance \label{sec:degradation_focusing}}

To show a more specific comparison, the focusing efficiency of the metasurface is calculated, considering both the angular and spectral dispersion of the field distribution at $F$ and $F'$. Three methods are compared, which are based on; the point dipole model with single-mode dispersion meta-atoms (Sec.~\ref{sec:angdip_lcmodel}), the point dipole model with multi-mode dispersion meta-atoms (Sec.~\ref{sec:angdip_fullwave}), and full-wave CST simulation (Sec.~\ref{sec:meas_sim}). A similar comparison of efficiency versus angular and spectral bandwidth has been reported recently in \cite{decker2019imaging} but only considering deflection efficiency from homogeneous metasurfaces. Here we consider the focusing efficiency of an inhomogeneous metasurface, which is calculated by the full-width half maximum (FWHM) of the field intensity at the focal spot $|E_y|^2$. The focusing efficiency here is a ratio between the integral of scattered field intensity considering three times of FWHM, to the integrated intensity of the incident wave $|E_{y,\mathrm{inc}}|^2$  \cite{arbabi2016miniature,arbabi2015subwavelength}. This can be formulated as
\begin{equation}
\mathrm{Focusing \; Efficiency} = \frac{\int_{-a}^a |E_{y}|^2 \, \mathrm{d}x}{\int_{-R}^R |E_{y,\mathrm{inc}}|^2 \, \mathrm{d}x} 
\end{equation}
For the incident wave, the integral limit $R$ refers to the radius of the metasurface, as illustrated in~\ref{fig:sample_meassetup}(a). For the focal point, the integral limit $a=1.5\times \mathrm{FWHM}$. 

\begin{figure*}[t]
    \centering
    \includegraphics[width=0.9\linewidth]{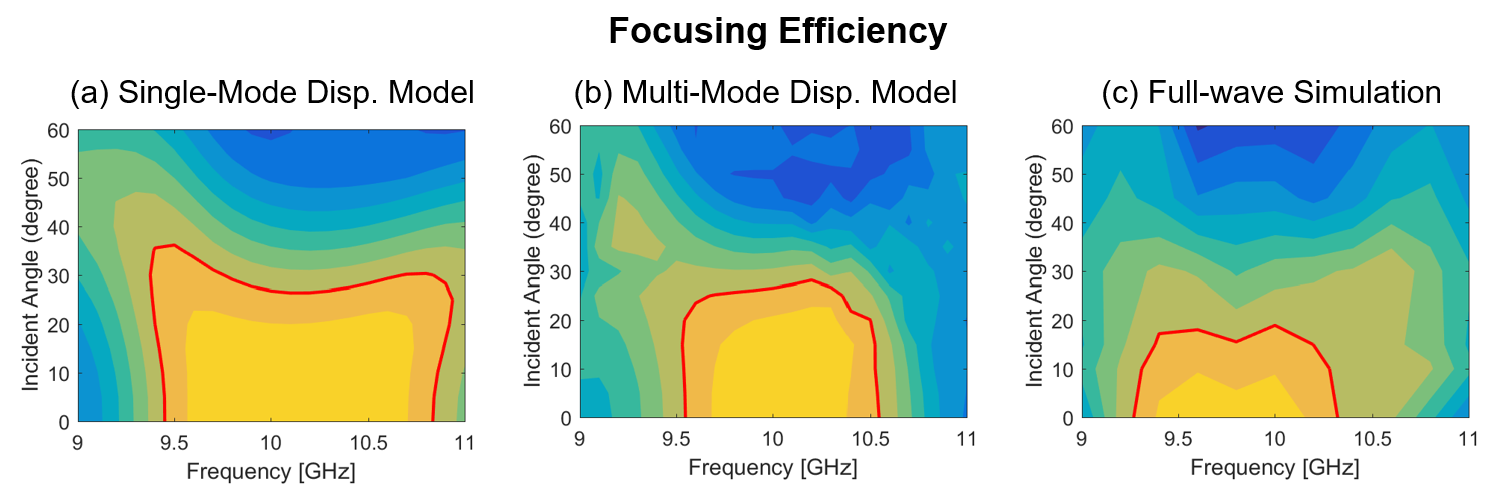}
    \caption{ Focusing efficiency over frequency and incident angles using; (a) point dipole calculation from single-mode dispersion model, (b) point dipole calculation from multi-mode dispersion model, and (c) full-wave simulation of the metasurface structure. Red lines indicate 80\% focusing efficiency.}
    \label{fig:focusing_eff}
\end{figure*}

As shown in (Fig.~\ref{fig:focusing_eff}(a)),  the focusing efficiency of the metasurface obtained by the single-mode dispersion model has the highest spectral and angular bandwidth. This focusing efficiency bandwidth does not agree well with the full-wave simulation result (Fig.~\ref{fig:focusing_eff}(c)), where the spectral and angular bandwidth are smaller, as shown by the red-lines indicating the 80\% efficiency limit. 

The focusing efficiency shown in Fig.~\ref{fig:focusing_eff}(b), is obtained from the point dipole calculation with multi-mode dispersion model, which captures the angular dispersion effect of additional resonances due to undesired dipoles and quadrupole modes. The result shows that it has lower spectral and angular bandwidth than using the single-mode dispersion model, and has a better agreement to the full-wave sumulation result, confirming that the additional resonances do degrade the metasurface focusing performance. These results suggest that beyond the monochromatic aberrations, the performance of metasurface at oblique incidence is also critically degraded by the angular dispersion of the meta-atoms. To realize metasurfaces working with wide incident angles, it is crucial to suppress this angular dispersion, which can be done by designing meta-atoms with minimum multi-polar mode coupling. 

\section{Conclusion}
We provide a thorough characterization of off-axis aberrations in a reflective focusing metasurface based on metallic-dielectric structures, commonly used at microwave and millimeter-wave frequencies. We consider a transverse electric (TE) plane-wave at various oblique incident angles, and  evaluate the metasurface focusing performance in broadband operation, providing a general picture of angular and spectral bandwidth of the metasurface. We characterize each meta-atom's angular dispersion and find that in addition to the designed resonances, additional resonances exist within the meta-atom, due to the dipole and quadrupole modes supported by the meta-atom geometry. The electric dipole resonances tend to increase in frequency with increasing incident angles (blue shift), while the quadrupole resonances tend to decrease with increasing incident angles  (red shift). This tendency contradicts the designed meta-atoms' operation, causing the phase and amplitude responses to strongly deviate at higher incident angles. The electric fields scattered by the metasurface are then calculated by using the point dipole model considering the angular and spectral dispersion of the meta-atoms. This allows calculation of the metasurface focusing efficiency, where the effect of meta-atom angular dispersion can be distinctly quantified. In contrast to previous studies, we show that the performance of the metasurface degrades beyond monochromatic aberrations predicted by the theory. The result suggest that to obtain broadband achromatic focusing metasurfaces in a wide range of incident angles, the meta-atoms' angular dispersion should be carefully taken into account. The numerical analysis provided in this chapter can also be used to characterize the transmissive Huygens' metasurface reported in \cite{fathnan2020achromatic}. At oblique incidence, additional resonances may appear as a result of higher order modes supported by the Huygens' meta-atoms, and a similar focusing degradation would be expected. 

\section*{Appendix}
\begin{figure}[h]
    \centering
    \includegraphics[width=0.8\linewidth]{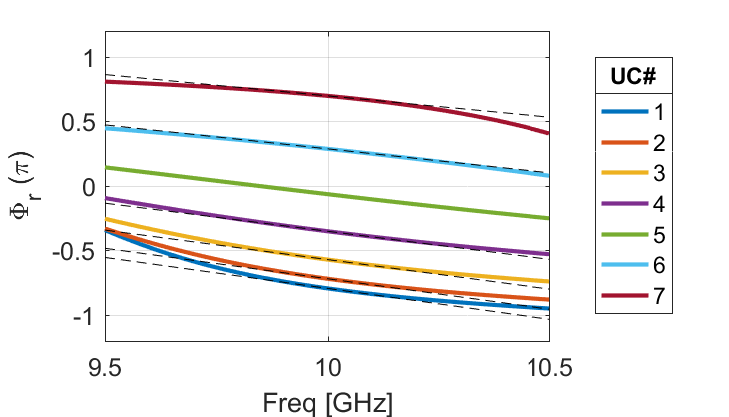}
    \caption{The achromatic metasurface phase profile over frequency from the design (dashed) and  LC approximation method (continuous), where the unit cell number are indicated.}
    \label{fig:ucphase}
\end{figure}

Meta-atom parameters for dog-bone and inverse dog-bone structures are presented in Table~\ref{tab:meta-atom-para}. The design achromatic metasurface phase profile over frequency is shown in Fig.~\ref{fig:ucphase}.

\begin{table}[h]
\centering
\caption{Dimensions of dog-bone (for UC\#1-6) and inverse dog-bone (for UC\#7) (in millimeters) and the corresponding phase ($\Phi_r$, in degree) at the center operating frequency 10\,Ghz.}
\label{tab:meta-atom-para}
\begin{tabular}{ccccc}
\hline 
UC   \# & ax   (bx) & ay   (by) & a1(b1) & Phase   at  10   GHz \\ \hline 
1       & 10.7      & 4.5       & 13.5   & -141.94              \\
2       & 11        & 4.5       & 13.5   & -121.10              \\
3       & 11.4      & 4.5       & 13.5   & -95.33               \\
4       & 12        & 4.5       & 13.5   & -64.07               \\
5       & 13        & 4.5       & 13     & -10.75               \\
6       & 12.5      & 4.2       & 13.5   & 43.45                \\
7     & 12.3      & 9.9       & 8.3    & 134.21               \\ \hline 
\multicolumn{5}{c}{UC \#1-6 a2=1, UC \#7 b2=1.35}         \\ \hline
\end{tabular}
\end{table}

\bibliographystyle{IEEEtran}
\bibliography{_biblio_file.bib}

\begin{thebibliography}{10}
\providecommand{\url}[1]{#1}
\csname url@samestyle\endcsname
\providecommand{\newblock}{\relax}
\providecommand{\bibinfo}[2]{#2}
\providecommand{\BIBentrySTDinterwordspacing}{\spaceskip=0pt\relax}
\providecommand{\BIBentryALTinterwordstretchfactor}{4}
\providecommand{\BIBentryALTinterwordspacing}{\spaceskip=\fontdimen2\font plus
\BIBentryALTinterwordstretchfactor\fontdimen3\font minus
\fontdimen4\font\relax}
\providecommand{\BIBforeignlanguage}[2]{{%
\expandafter\ifx\csname l@#1\endcsname\relax
\typeout{** WARNING: IEEEtran.bst: No hyphenation pattern has been}%
\typeout{** loaded for the language `#1'. Using the pattern for}%
\typeout{** the default language instead.}%
\else
\language=\csname l@#1\endcsname
\fi
#2}}
\providecommand{\BIBdecl}{\relax}
\BIBdecl

\bibitem{fathnan2020achromatic}
A.~A. Fathnan, M.~Liu, and D.~A. Powell, ``Achromatic huygensâ€™ metalenses
with deeply subwavelength thickness,'' \emph{Advanced Optical Materials}, p.
2000754, 2020.

\bibitem{tsilipakos2020squeezing}
O.~Tsilipakos, M.~Kafesaki, E.~N. Economou, C.~M. Soukoulis, and T.~Koschny,
``Squeezing a prism into a surface: Emulating bulk optics with achromatic
metasurfaces,'' \emph{Advanced Optical Materials}, p. 2000942, 2020.

\bibitem{li2014diffraction}
Y.~B. Li, B.~G. Cai, X.~Wan, and T.~J. Cui, ``Diffraction-free surface waves by
metasurfaces,'' \emph{Optics letters}, vol.~39, no.~20, pp. 5888--5891, 2014.

\bibitem{hadad2015space}
Y.~Hadad, D.~L. Sounas, and A.~Alu, ``Space-time gradient metasurfaces,''
\emph{Physical Review B}, vol.~92, no.~10, p. 100304, 2015.

\bibitem{he2018waveguide}
Y.~He, Y.~Li, L.~Zhu, H.~Bagci, D.~Erricolo, and P.-Y. Chen, ``Waveguide
dispersion tailoring by using embedded impedance surfaces,'' \emph{Physical
Review Applied}, vol.~10, no.~6, p. 064024, 2018.

\bibitem{chen2019design}
M.~Chen, A.~Epstein, and G.~V. Eleftheriades, ``Design and experimental
verification of a passive huygensâ€™ metasurface lens for gain enhancement of
frequency-scanning slotted-waveguide antennas,'' \emph{IEEE Transactions on
Antennas and Propagation}, vol.~67, no.~7, pp. 4678--4692, 2019.

\bibitem{yu2018design}
S.~Yu, H.~Liu, and L.~Li, ``Design of near-field focused metasurface for
high-efficient wireless power transfer with multifocus characteristics,''
\emph{IEEE Transactions on Industrial Electronics}, vol.~66, no.~5, pp.
3993--4002, 2018.

\bibitem{smith2017analysis}
D.~R. Smith, V.~R. Gowda, O.~Yurduseven, S.~Larouche, G.~Lipworth, Y.~Urzhumov,
and M.~S. Reynolds, ``An analysis of beamed wireless power transfer in the
fresnel zone using a dynamic, metasurface aperture,'' \emph{Journal of
Applied Physics}, vol. 121, no.~1, p. 014901, 2017.

\bibitem{schoenlinner2002wide}
B.~Schoenlinner, X.~Wu, J.~P. Ebling, G.~V. Eleftheriades, and G.~M. Rebeiz,
``Wide-scan spherical-lens antennas for automotive radars,'' \emph{IEEE
Transactions on microwave theory and techniques}, vol.~50, no.~9, pp.
2166--2175, 2002.

\bibitem{saleem2017lens}
M.~K. Saleem, H.~Vettikaladi, M.~A. Alkanhal, and M.~Himdi, ``Lens antenna for
wide angle beam scanning at 79 ghz for automotive short range radar
applications,'' \emph{IEEE Transactions on Antennas and Propagation},
vol.~65, no.~4, pp. 2041--2046, 2017.

\bibitem{wyant1992basic}
J.~C. Wyant and K.~Creath, ``Basic wavefront aberration theory for optical
metrology,'' \emph{Applied optics and optical engineering}, vol.~11, no. part
2, pp. 28--39, 1992.

\bibitem{young1972zone}
M.~Young, ``Zone plates and their aberrations,'' \emph{JOSA}, vol.~62, no.~8,
pp. 972--976, 1972.

\bibitem{delano1974primary}
E.~Delano, ``Primary aberrations of fresnel lenses,'' \emph{JOSA}, vol.~64,
no.~4, pp. 459--468, 1974.

\bibitem{liang2019high}
H.~Liang, A.~Martins, B.-H.~V. Borges, J.~Zhou, E.~R. Martins, J.~Li, and T.~F.
Krauss, ``High performance metalenses: numerical aperture, aberrations,
chromaticity, and trade-offs,'' \emph{Optica}, vol.~6, no.~12, pp.
1461--1470, 2019.

\bibitem{decker2019imaging}
M.~Decker, W.~T. Chen, T.~Nobis, A.~Y. Zhu, M.~Khorasaninejad, Z.~Bharwani,
F.~Capasso, and J.~Petschulat, ``Imaging performance of
polarization-insensitive metalenses,'' \emph{ACS Photonics}, vol.~6, no.~6,
pp. 1493--1499, 2019.

\bibitem{kalvach2016aberration}
A.~Kalvach and Z.~Szab{\'o}, ``Aberration-free flat lens design for a wide
range of incident angles,'' \emph{JOSA B}, vol.~33, no.~2, pp. A66--A71,
2016.

\bibitem{aieta2013aberrations}
F.~Aieta, P.~Genevet, M.~Kats, and F.~Capasso, ``Aberrations of flat lenses and
aplanatic metasurfaces,'' \emph{Optics express}, vol.~21, no.~25, pp.
31\,530--31\,539, 2013.

\bibitem{arbabi2016miniature}
A.~Arbabi, E.~Arbabi, S.~M. Kamali, Y.~Horie, S.~Han, and A.~Faraon,
``Miniature optical planar camera based on a wide-angle metasurface doublet
corrected for monochromatic aberrations,'' \emph{Nature communications},
vol.~7, no.~1, pp. 1--9, 2016.

\bibitem{minovich2010tilted}
A.~Minovich, D.~N. Neshev, D.~A. Powell, I.~V. Shadrivov, M.~Lapine,
I.~McKerracher, H.~T. Hattori, H.~H. Tan, C.~Jagadish, and Y.~S. Kivshar,
``Tilted response of fishnet metamaterials at near-infrared optical
wavelengths,'' \emph{Physical Review B}, vol.~81, no.~11, p. 115109, 2010.

\bibitem{albooyeh2014resonant}
M.~Albooyeh, S.~Kruk, C.~Menzel, C.~Helgert, M.~Kroll, A.~Krysinski, M.~Decker,
D.~N. Neshev, T.~Pertsch, C.~Etrich \emph{et~al.}, ``Resonant metasurfaces at
oblique incidence: interplay of order and disorder,'' \emph{Scientific
reports}, vol.~4, no.~1, pp. 1--7, 2014.

\bibitem{qiu2018angular}
M.~Qiu, M.~Jia, S.~Ma, S.~Sun, Q.~He, and L.~Zhou, ``Angular dispersions in
terahertz metasurfaces: physics and applications,'' \emph{Physical Review
Applied}, vol.~9, no.~5, p. 054050, 2018.

\bibitem{zhang2020controlling}
X.~Zhang, Q.~Li, F.~Liu, M.~Qiu, S.~Sun, Q.~He, and L.~Zhou, ``Controlling
angular dispersions in optical metasurfaces,'' \emph{Light: Science \&
Applications}, vol.~9, no.~1, pp. 1--12, 2020.

\bibitem{yazdi2017analysis}
M.~Yazdi and M.~Albooyeh, ``Analysis of metasurfaces at oblique incidence,''
\emph{IEEE Transactions on Antennas and Propagation}, vol.~65, no.~5, pp.
2397--2404, 2017.

\bibitem{fathnan2020bandwidth}
A.~A. Fathnan, A.~Olk, and D.~Powell, ``Bandwidth limit and synthesis approach
for single resonance ultrathin metasurfaces,'' \emph{Journal of Physics D:
Applied Physics}, 2020.

\bibitem{zhang2016high}
S.~Zhang, M.-H. Kim, F.~Aieta, A.~She, T.~Mansuripur, I.~Gabay,
M.~Khorasaninejad, D.~Rousso, X.~Wang, M.~Troccoli \emph{et~al.}, ``High
efficiency near diffraction-limited mid-infrared flat lenses based on
metasurface reflectarrays,'' \emph{Optics express}, vol.~24, no.~16, pp.
18\,024--18\,034, 2016.

\bibitem{olk2020huygens}
A.~E. Olk and D.~A. Powell, ``Huygens metasurface lens for w-band switched beam
antenna applications,'' \emph{IEEE Open Journal of Antennas and Propagation},
vol.~1, pp. 290--299, 2020.

\bibitem{akram2020bi}
M.~R. Akram, C.~He, and W.~Zhu, ``Bi-layer metasurface based on huygensâ€™
principle for high gain antenna applications,'' \emph{Optics Express},
vol.~28, no.~11, pp. 15\,844--15\,854, 2020.

\bibitem{abbasi2019millimeter}
M.~I. Abbasi, M.~H. Dahri, M.~H. Jamaluddin, N.~Seman, M.~R. Kamarudin, and
N.~H. Sulaiman, ``Millimeter wave beam steering reflectarray antenna based on
mechanical rotation of array,'' \emph{IEEE Access}, vol.~7, pp.
145\,685--145\,691, 2019.

\bibitem{kiris2019reflectarray}
O.~Kiris, K.~Topalli, and M.~Unlu, ``A reflectarray antenna using hexagonal
lattice with enhanced beam steering capability,'' \emph{IEEE Access}, vol.~7,
pp. 45\,526--45\,532, 2019.

\bibitem{li1981focal}
Y.~Li and E.~Wolf, ``Focal shifts in diffracted converging spherical waves,''
\emph{Optics communications}, vol.~39, no.~4, pp. 211--215, 1981.

\bibitem{yang2014efficient}
Q.~Yang, J.~Gu, D.~Wang, X.~Zhang, Z.~Tian, C.~Ouyang, R.~Singh, J.~Han, and
W.~Zhang, ``Efficient flat metasurface lens for terahertz imaging,''
\emph{Optics express}, vol.~22, no.~21, pp. 25\,931--25\,939, 2014.

\bibitem{born1999principles}
M.~Born and E.~Wolf, ``Principles of optics, 7th (expanded) edition,''
\emph{United Kingdom: Press Syndicate of the University of Cambridge}, vol.
461, 1999.

\bibitem{ginn2010monochromatic}
J.~Ginn, J.~Alda, J.~A. G{\'o}mez-Pedrero, and G.~Boreman, ``Monochromatic
aberrations in resonant optical elements applied to a focusing multilevel
reflectarray,'' \emph{Optics Express}, vol.~18, no.~11, pp. 10\,931--10\,940,
2010.

\bibitem{liu2017high}
Z.~Liu, S.~Du, A.~Cui, Z.~Li, Y.~Fan, S.~Chen, W.~Li, J.~Li, and C.~Gu,
``High-quality-factor mid-infrared toroidal excitation in folded 3d
metamaterials,'' \emph{Advanced Materials}, vol.~29, no.~17, p. 1606298,
2017.

\bibitem{liu2010planar}
N.~Liu, T.~Weiss, M.~Mesch, L.~Langguth, U.~Eigenthaler, M.~Hirscher,
C.~Sonnichsen, and H.~Giessen, ``Planar metamaterial analogue of
electromagnetically induced transparency for plasmonic sensing,'' \emph{Nano
letters}, vol.~10, no.~4, pp. 1103--1107, 2010.

\bibitem{tao2008highly}
H.~Tao, C.~Bingham, A.~Strikwerda, D.~Pilon, D.~Shrekenhamer, N.~Landy, K.~Fan,
X.~Zhang, W.~Padilla, and R.~Averitt, ``Highly flexible wide angle of
incidence terahertz metamaterial absorber: Design, fabrication, and
characterization,'' \emph{physical review B}, vol.~78, no.~24, p. 241103,
2008.

\bibitem{tao2010dual}
H.~Tao, C.~Bingham, D.~Pilon, K.~Fan, A.~Strikwerda, D.~Shrekenhamer,
W.~Padilla, X.~Zhang, and R.~Averitt, ``A dual band terahertz metamaterial
absorber,'' \emph{Journal of physics D: Applied physics}, vol.~43, no.~22, p.
225102, 2010.

\bibitem{zhang2013multispectral}
B.~Zhang, J.~Hendrickson, and J.~Guo, ``Multispectral near-perfect metamaterial
absorbers using spatially multiplexed plasmon resonance metal square
structures,'' \emph{JOSA B}, vol.~30, no.~3, pp. 656--662, 2013.

\bibitem{he2011dual}
X.-J. He, Y.~Wang, J.~Wang, T.~Gui, and Q.~Wu, ``Dual-band terahertz
metamaterial absorber with polarization insensitivity and wide incident
angle,'' \emph{Progress In Electromagnetics Research}, vol. 115, pp.
381--397, 2011.

\bibitem{arbabi2015subwavelength}
A.~Arbabi, Y.~Horie, A.~J. Ball, M.~Bagheri, and A.~Faraon,
``Subwavelength-thick lenses with high numerical apertures and large
efficiency based on high-contrast transmitarrays,'' \emph{Nature
communications}, vol.~6, no.~1, pp. 1--6, 2015.

\end{thebibliography}

\end{document}